\documentclass[a4paper,fleqn,usenatbib]{mnras}

\usepackage[T1]{fontenc}
\usepackage{ae,aecompl}

\usepackage{graphicx}	% Including figure files
\usepackage{amsmath}	% Advanced maths commands
\usepackage{amssymb}	% Extra maths symbols

\def\apj{ApJ}
\def\apjs{ApJS}
\def\mnras{MNRAS}
\def\nat{Nature}
\def\apjl{ApJL}
\def\aap{A\&A}
\def\aj{AJ}
\def\aplett{ApL}

\def\asec{\ifmmode ^{\prime\prime}\else$^{\prime\prime}$\fi}

\def\it{\sl}
\def\degs{\ifmmode ^{\circ}\else$^{\circ}$\fi}
\def\amin{\ifmmode ^{\prime}\else$^{\prime}$\fi}
\def\asec{\ifmmode ^{\prime\prime}\else$^{\prime\prime}$\fi}
\def\farcs{\hbox{$.\!\!^{\prime\prime}$}}  % Fractions of arcseconds
\def\degs{\ifmmode ^{\circ}\else$^{\circ}$\fi}
\def\amin{\ifmmode ^{\prime}\else$^{\prime}$\fi}
\def\farcm{\hbox{$.\mkern-4mu^\prime$}}
\def\eqalign#1{\null\,\vcenter{\openup1\jot \m@th
   \ialign{\strut\hfil$\displaystyle{##}$&$\displaystyle{{}##}$\hfil
   \crcr#1\crcr}}\,}
\title[Observations of PSR J1357$-$6429 at 2.1 GHz]{Observations of PSR J1357$-$6429 at 2.1 GHz with the  Australia Telescope
Compact Array}

\author[A. Kirichenko et al.]{A.~Kirichenko,$^{1,2}$\thanks{E-mail: aida.astro@mail.ioffe.ru (AK)}
Yu.~Shibanov,$^{1,2}$
P.~Shternin,$^{1,2}$
S.~Johnston,$^3$
M.~A.~Voronkov,$^{3,4,5}$
\newauthor
A.~Danilenko,$^{1}$ 
D.~Barsukov,$^{1,2}$
D.~Lai$^{6}$
and D.~Zyuzin$^{1}$
\\
$^{1}$Ioffe Institute, 26 Politekhnicheskaya st., St. Petersburg 194021, Russia\\
$^{2}$Peter the Great St. Petersburg Polytechnic University, 29 Politekhnicheskaya st., St. Petersburg 195251, Russia\\
$^{3}$CSIRO Astronomy and Space Science, Australia Telescope National Facility, PO Box 76, Epping, NSW 1710, Australia\\
$^{4}$Astro Space Centre, Profsouznaya st. 84/32, 117997 Moscow, Russia\\
$^{5}$School of Mathematics and Physics, University of Tasmania, GPO Box 252-37, Hobart, Tasmania 7000, Australia\\
$^{6}$Department of Astronomy, Cornell University, Ithaca, NY 14853, USA\\
}

\date{Accepted XXX. Received YYY; in original form ZZZ}

\pubyear{2015}

\begin{document}
\label{firstpage}
\pagerange{\pageref{firstpage}--\pageref{lastpage}}
\maketitle

\begin{abstract}
PSR J1357$-$6429 is a young and energetic radio pulsar detected in X-rays and $\gamma$-rays. 
It powers a compact pulsar wind
nebula with a jet visible in X-rays and a large scale plerion detected in  X-ray and TeV ranges.
Previous multiwavelength studies suggested that the pulsar has a significant proper motion of about 
180 mas yr$^{-1}$ implying
an extremely high transverse velocity of about 2000 km~s$^{-1}$.
In order to verify that, we performed radio-interferometric observations of PSR J1357$-$6429 
with the the Australia Telescope Compact Array (ATCA) %
in the 2.1 GHz band. We detected the pulsar 
with a mean flux density of $212\pm5$ $\mu$Jy and  
obtained the most accurate  pulsar position, RA~= 13:57:02.525(14) and Dec~= $-$64:29:29.89(15).
 Using  the new  and archival ATCA data, we did not find any proper motion and estimated its 90 per cent upper limit
$\mu < 106$ mas yr$^{-1}$.
The pulsar shows a highly polarised single pulse, as it was earlier observed at 1.4 GHz. 
Spectral analysis revealed a shallow spectral
index $\alpha_{\nu}$ = $0.5 \pm 0.1$. 
Based on our new radio position of the pulsar, we disclaim its optical counterpart candidate reported before.
\end{abstract}


%%%%%%%%%%%%%%%%%%%%%%%%%%%%%%%%%%%%%%%%%%%%%%%%%%

%%%%%%%%%%%%%%%%% BODY OF PAPER %%%%%%%%%%%%%%%%%%

\section{Introduction}
\label{Sect:intro}
Born in supernova explosions, neutron stars (NSs) typically obtain velocities orders of magnitude greater
than those of their stellar progenitors.
According to statistical analyses \citep[see][]{hobbs},
the mean three-dimensional velocities of NSs mostly derived from radio data are
about 400 km~s$^{-1}$. The largest firmly established pulsar transverse velocity 
of 1080~$\pm$~100 km~s$^{-1}$, which has been determined
by direct proper motion and parallax measurements with the VLBA, belongs to PSR B1508+55 \citep{chatterjee}.

Although an initial kick which occurs during supernova explosions is generally accepted as the reason for high
velocities, the origin of the kick is currently unclear. \citet{lai}
discuss several classes of kick mechanisms, but it is not clear whether any of them can fully explain 
the fastest moving neutron stars
\citep[e.g.,][]{chatterjee}. Therefore, new detections of high-velocity pulsars are needed to 
put additional constraints on models used in the latest supernova explosion simulations
\citep[see][]{janka}. The velocity measurements might be also
potentially useful to reveal possible relationship between the velocity and other
parameters of NSs \citep{lai}.

PSR J1357$-$6429 is a young (characteristic age 7.3 kyr) and energetic 
(spin-down luminosity $\dot{E}$ = 3.1$\times10^{36}$ ergs~s$^{-1}$)
radio pulsar with a period of 166 ms \citep{camilo}. The pulse profile and polarisation 
were studied with the Parkes telescope at 1.4 GHz by \citet{camilo, johnston, lemo} and \citet{rookyard2015MNRAS}.   
The pulsar field was observed in X-rays, where the pulsar counterpart and a compact tail-like
pulsar wind nebula (PWN), implying a noticeable pulsar proper motion, were found
\citep{espo, zavlin}. Thorough follow-up high-energy studies
have revealed X-ray and $\gamma$-ray pulsations
with the pulsar period \citep{lemo, chang}.
A pulsar plerion extended to a few tens of arcminutes was detected in %the radio, 
X-ray and TeV ranges by \citet{abram}. They also argued that the extended radio emission from 
the supernova remnant (SNR)  candidate G309.8$-$2.6 \citep{duncan} is rather  the plerion counterpart.  

\citet{danilenko} reported the detection of a faint pulsar optical counterpart candidate
with the Very Large Telescope (VLT) whose position was in an agreement with PSR J1357$-$6429 X-ray position.
However, a significant offset (1\farcs54~$\pm$~0\farcs32) of the  candidate position
from the J1357$-$6429 radio-interferometric coordinates measured 8.7~yr earlier by \citet{camilo} with the Australia Telescope
Compact Array (ATCA), rises some doubts about the pulsar nature of the candidate and/or
suggests that the pulsar has an extremely  high proper motion.

The distance to the pulsar of $\sim$2.5 kpc was estimated from
its dispersion measure $DM=128.5$ pc cm$^{-3}$ by \citet{camilo}. 
\citet[][]{danilenko} performed an independent distance analysis
comparing the interstellar extinction--distance relation along the pulsar line of sight 
and the absorbing column density obtained from the X-ray spectral analysis.
The resulting distance range of 2.0--2.5 kpc supports the DM distance estimate.
The distance range and the candidate offset imply the pulsar transverse velocity to be between
1300 km~s$^{-1}$ and 2500 km~s$^{-1}$, which is 
higher than the largest NS velocity precisely measured
so far \citep{chatterjee}. A similar velocity range was estimated from the comparison of 
the pulsar X-ray and radio-interferometric positions \citep{mignani}.

Aiming to check whether the pulsar velocity is indeed that high, we performed
new radio-interferometric observations with the ATCA to obtain a precise pulsar position for another
epoch. Another reason for the observations was  
to extend the radio studies of the pulsar itself, which was investigated only at 1.4 GHz, to a higher frequency range.
The details of observations and data reduction are described
in Sect.~\ref{sec2}. Our results and reanalysis of the archival data are presented in Sect.~\ref{sec3}
and summarised in Sect.~\ref{sec4}.

\begin{figure*}
\setlength{\unitlength}{1mm}
  \begin{center}
    \begin{picture}(145,75)(0,0)
 \put  (75,0.) {\includegraphics[scale=0.5, clip]{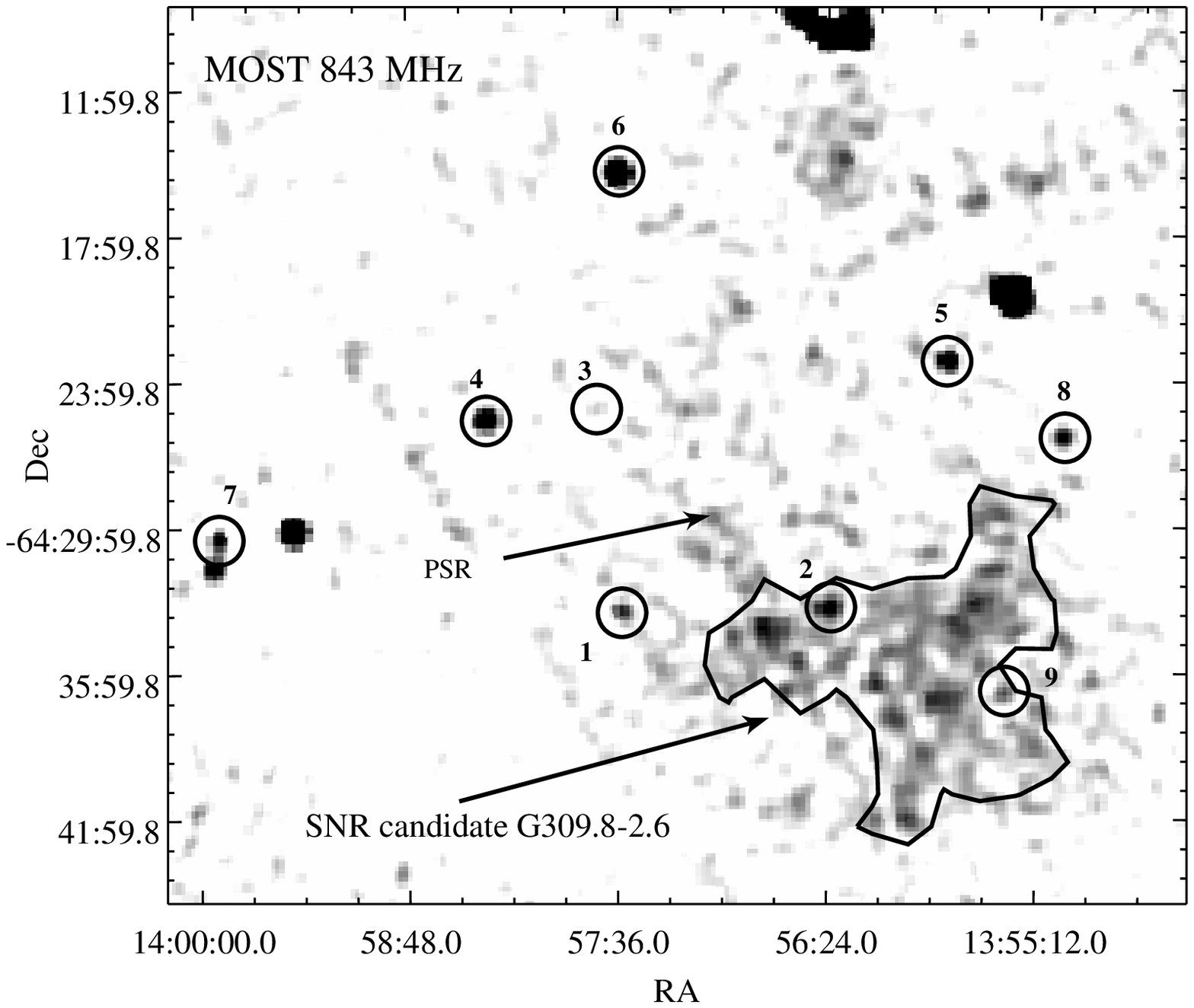}}
 \put  (-16,0.){\includegraphics[scale=0.49, clip]{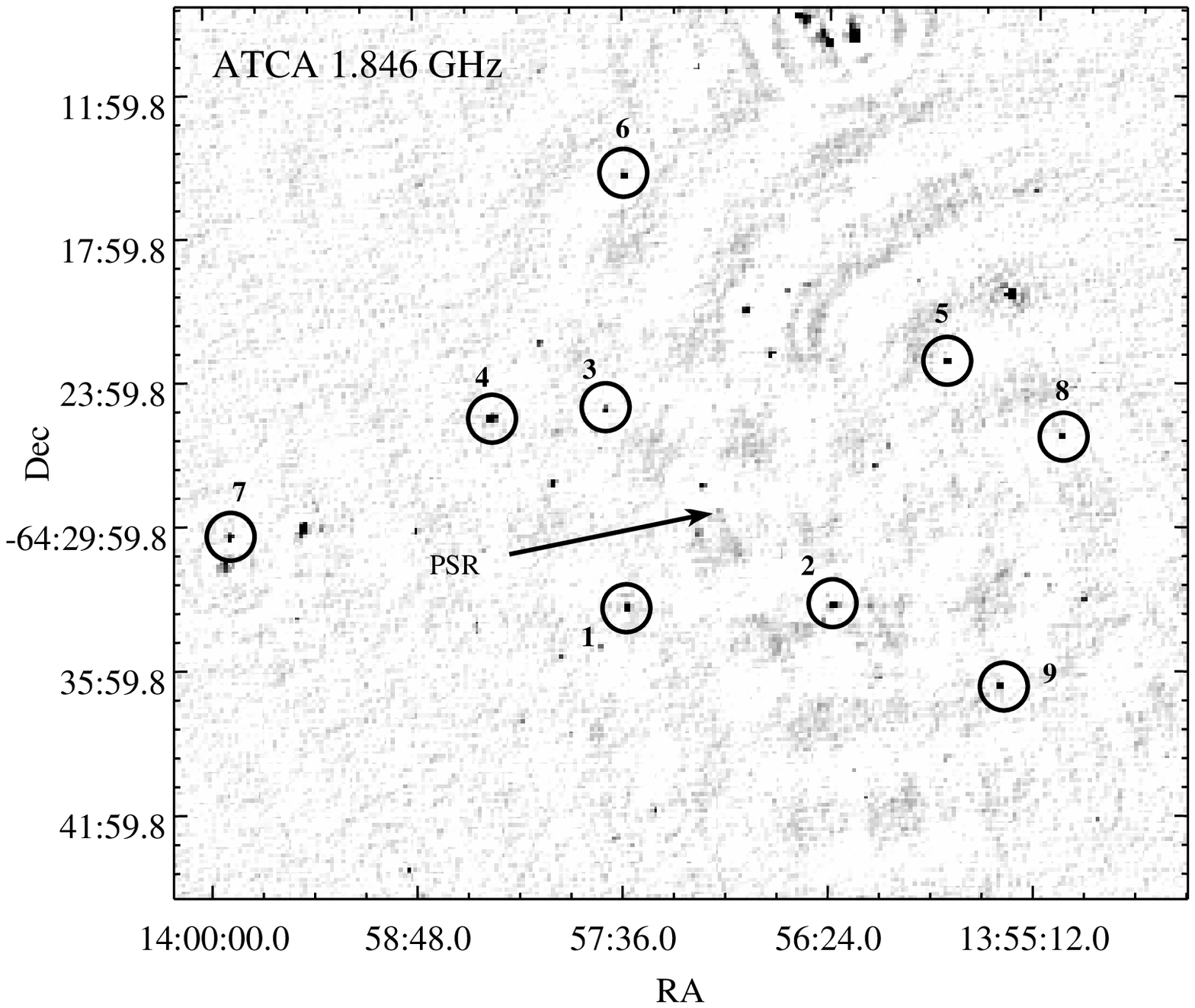}}
 \end{picture}
  \end{center}
  \caption{$50\arcmin\times50\arcmin$ field fragment centred at 
  the PSR J1357$-$6429 as it is visible with the ATCA
  ({\sl left}) and the MOST ({\sl right}) 
  at 1.846 GHz and 843 MHz, respectively. 
  The positions of the pulsar
  and the extended emission from the SNR candidate G309.8$-$2.6 surrounded by the contour
  are indicated by arrows.
  Numbered circles mark nine bright point-like objects used for the relative astrometry of
  the ATCA images (see text for details).
  }
\label{fig:FOV}
\end{figure*}

\section{Observations and data reduction}
\label{sec2}

The observations of the pulsar field were carried out with the ATCA on 2013 June 3.
The observing session started at UT 4:30 and lasted for 9.5 hours. 
The main goal of the observations was to measure a precise position
of the source. Therefore,  
the observations were performed with the
6C array configuration, which has the maximum baseline of nearly 6 km. We used the pulsar binning capability of
the Compact Array Broadband Backend (CABB), which allowed us to perform high time resolution observations \citep{wilson}.
The binning mode split the 166 ms
pulsar period into 32 independent rotational phase bins.    
The observations were carried out in the 16 cm band centred at 2.102 GHz. The
total bandwidth of 2.048 GHz was split into 512 spectral channels providing 4 MHz spectral 
resolution in the 1.078--3.126 GHz range.
PKS B1934$-$638 was observed at the beginning of the session as a primary standard for  
the flux density scale and bandpass calibrations.
To account for gain and phase instabilities, we observed two nearby secondary calibrators 
1329$-$665 and 1325$-$55 in a ten-minute loop with the pulsar,
where each calibrator was observed for about two minutes. 
Two calibrators were used to estimate the systematic errors on  the position reference frame.
Standard data reduction including Radio Frequency Interference (RFI) 
flagging and calibration was performed with {\tt MIRIAD} package \citep{miriad},
and {\tt Karma} \citep{gooch} tools were used for data visualisation.

The data were split
into four 512 MHz sub-bands with central frequencies of 1.334 GHz, 1.846 GHz, 2.358 GHz and 2.870 GHz.
The secondary calibration was then performed separately for each bandwidth partition.

The data were imaged using the MIRIAD {\sl invert} task with the ``robust'' parameter set to zero. 
It provides a tradeoff between better
signal-to-noise ratio ($S/N$) and stronger sidelobe suppression in the visibility weighting scheme.
The deconvolution process was performed with the {\sl mfclean} routine, which accounts for spectral variations across the bandwidth.
For the purposes of absolute astrometry, we used the resulting clean images without further calibration.

Phase self-calibration
was performed on the data to improve the image quality. After the first iteration of {\sl mfclean}, model
components with flux densities $\ga 1 $mJy beam$^{-1}$ were used for the initial multi-frequency phase
self-calibration, which considerably decreased residual phase errors. Then the weaker sources found on clean images were included 
in the self-calibration model. 
As a result, after several clean-self-calibration cycles, the sidelobes of field sources became
comparable with the thermal noise of $\approx50$ $\mu$Jy beam$^{-1}$ for the 1.334 GHz subband 
and $\approx25$--$30$ $\mu$Jy beam$^{-1}$ for the three other sub-bands. 

For the lower (1.334 GHz) and higher
(2.870 GHz) frequency sub-bands, the synthesised beam size was
$8\farcs3\times5\farcs1$ and
$4\farcs5\times2\farcs6$, respectively, with the position angle $PA \approx -24{^\circ}$.
The ATCA primary beam full width at half maximum (FWHM) was $42\arcmin$ for the lowest frequency, 
decreasing to $15\arcmin$ for the highest frequency.
The selected image size for all sub-bands was $\sim160\arcmin\times160\arcmin$.

To confidently measure the pulsar proper motion, we performed an independent analysis of 
the available archival ATCA data\footnote{http://atoa.atnf.csiro.au, project CX001.}. The set was
obtained on 2000 August 29 in the 6A configuration with the baselines close to those of 6C
but in 1.376 and 2.496 GHz bands with 128 MHz bandwidths each split in 14 spectral channels \citep{camilo}\footnote{The
2.496 GHz data were not reported by \citet{camilo}.}.
The data processing and imaging were similar to those applied to our own set,
except that no splitting was performed and the gain calibration
was done with the single observed calibrator 1329$-$665. The synthesised beam size was 
$12\farcs7\times5\farcs6$ with $PA \approx$~48\degs\ and $7\farcs3\times2\farcs2$ with $PA\approx$~40\degs\ 
for 1.376 and 2.496 GHz, respectively.
The imaging procedure for the 1.376 GHz band revealed a point-like
source surrounded by concentric sidelobe-like rings  which were difficult to clean off. 
The source is located close
to the phase centre and the pulsar, and was not identified in the 2013 images.
Thorough data inspection showed that it only
appears in channels 13 and 14 near the low frequency sideband and thus is likely spurious.
Given its proximity to the pulsar, we used only channels 1--12 to eliminate the artifact and its sidelobes
which could affect the pulsar position and flux measurements.

\section{Results}
\label{sec3}

\subsection{The pulsar field}

In Fig.~\ref{fig:FOV}, we show
the self-calibrated image of the pulsar field in the 1.846 GHz sub-band.
For comparison, we also present the Molonglo Observatory Synthesis Telescope (MOST) image at 843 MHz with
a restoring beam size of $\approx$~104\arcsec\ \citep{murphy}.
Many MOST point-like objects 
have their firm counterparts in the ATCA image and vice versa.
The pulsar is detected with the ATCA as a point-like source at a 10$\sigma$ significance in this sub-band  
and is also visible in the MOST image.

Due to significantly higher spatial resolution of the ATCA, several 
MOST sources are resolved as groups of separate objects. For instance, a bright extended object near the
north edge of the MOST image  is actually a blend of three compact objects.
At the same time, insufficient uv-coverage and the long baseline configuration make the interferometer
insensitive to large extended structures.  That is why the extended emission from the SNR candidate
G309.8$-$2.6 \citep{duncan} and other extended structures visible in the MOST image 
cannot be identified in the ATCA image.

The compact NE-SW tail-like PWN structure of J1357$-$6429 visible in X-rays \citep{chang} 
is not seen in the radio.

\subsection{The astrometry and pulsar proper motion}

To constrain the pulsar proper motion, we used two astrometric methods. 
The first one, which we hereafter refer to as ``absolute astrometry'', 
allows to measure source coordinates relative to phase calibrators, 
whose positions are known with a high accuracy, for instance, from the VLBI measurements.  
The second approach, which we entitle ``relative astrometry'', is based on 
measurements of the target position shift between observational epochs relative to other sources in the field.

In all cases, to measure the pulsar positions, 
we used on-pulse data
obtained from the calibrated visibilities.
Using the {\tt MIRIAD} {\sl psrfix} routine, the pulse-phase bins were phase-adjusted as a function of frequency
channel accounting for the known DM of 128.5~pc~cm$^{-3}$ and the pulsar period of 166~ms.
After that, a mean off-pulse baseline value was subtracted with the {\sl psrbl} tool.
This considerably decreased the pulsar contamination by backgrounds.

\subsubsection{Absolute astrometry}

We obtained eight positions of the pulsar in the 2013 images, for
each of the four spectral sub-bands
in both secondary calibrations.
They were determined
with the {\tt MIRIAD} task {\sl imfit} using the object type ``point''.
The position uncertainty derived with this task 
slightly depends on the size of a region around the pulsar where the fit is performed.
We used a region of 
about twice the synthesised beam size in each sub-band. We checked that the resulting positional errors were comparable 
to the size of the beam, 
divided by twice the (S/N). The S/N was $\approx$ 11.2, 9.1, 
10.8 and 6.4 for the 1.334, 1.846, 2.358 and 2.780 GHz sub-bands, respectively.
To provide better position measurements, 
the pixel grid on the on-pulse images was adjusted to place 
the  pulsar exactly in the pixel centre.
The positions measured in each sub-band were found to be consistent 
within uncertainties, indicating
that no correction for systematic  errors between the sub-bands is needed \citep{deller}.

\begin{figure}
  \begin{center}
    \includegraphics[scale=0.3,clip]{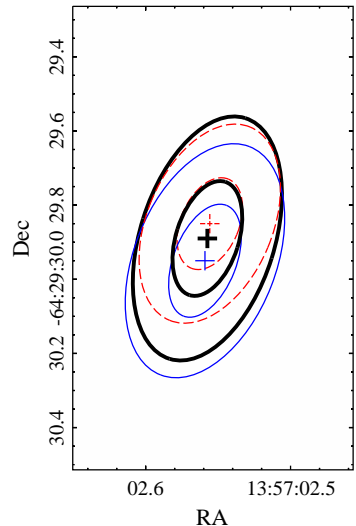}
  \end{center}
  \caption{Pulsar position uncertainty ellipses measured at 40 and 90 per cent confidence levels in 2.1
  GHz band. Thin solid and dashed line crosses and ellipses correspond to 
  the calibrations obtained with 1329$-$665 and 1325$-$55 standards, respectively.
  The bold cross and solid bold line ellipses
  are the derived weighted mean positions.} 
\label{fig:ellipses}
\end{figure}

The mean pulsar positions for 1325$-$665 and 1329$-$55 calibrations weighted over the four sub-bands
are RA~= 13:57:02.526(15), Dec~= $-$64:29:29.95(15) and  RA~= 13:57:02.524(14),  Dec~= $-$64:29:29.85(12)\footnote{Herein,
the numbers in brackets are 1$\sigma$ uncertainties referring to the last significant
digits  quoted, the equinox is J2000.0.},
respectively. The best-fit positions and the positional
error ellipses are shown in Fig.~\ref{fig:ellipses} with crosses and inner thin solid and dashed ellipses for
1325$-$65 and 1329$-$55 calibrations, respectively. The positional error ellipse  projected
on the 1$\sigma$ coordinate uncertainty represents only 40 per cent 2D confidence
region \citep[e.g.,][]{2002nr}. For completeness, we also show  
90 per cent confidence ellipses, which are by a factor
$\approx 2.14$ larger in size. The two measured positions show $\approx0\farcs12$ systematic offset
in Dec. We checked that it is consistent with
the shift obtained after cross-calibrating the 1325$-$665 and 1329$-$55 standards 
and measuring their positions. 
According to Fig.~\ref{fig:ellipses}, the estimated systematic error along the
synthesized beam major principle axis is comparable with the formal statistical 1$\sigma$ uncertainty.
Therefore, the two source positions were  weighted by the respective covariance matrices, in
order to account for correlations between the errors in RA and Dec, and combined. The resulting mean
position with 40 per cent and 90 per cent uncertainties are shown in Fig.~\ref{fig:ellipses} with the bold cross and
ellipses. The uncertainties were obtained by adding the estimated systematic covariance
matrix to the weighted mean statistical covariance matrix.
The final coordinates are RA~= 13:57:02.525(14) and Dec~= $-$64:29:29.89(15), where 1$\sigma$ errors  correspond 
to the inner bold
ellipse in Fig.~\ref{fig:ellipses}.

The derived position appears to be different from RA~= 13:57:02.43(2) and Dec~= $-$64:29:30.2(1) provided by \citet{camilo}
based on the 2000 ATCA 1.376 GHz observations
implying that the pulsar has moved between 2000 and 2013 epochs. 

To check the pulsar shift significance, we remeasured the pulsar position on the 1.376 GHz on-pulse 
image of the 2000 data where the pulsar is detected with S/N$\approx$13.
At 2.496 GHz,
the pulsar is found at a significantly lower S/N, making the position measurements less accurate.
The derived 1.376 GHz coordinates are RA~= 13:57:02.546(76) and Dec~= $-$64:29:29.64(55)\footnote{The systematics cannot 
be accounted here since only one secondary calibrator was observed.}.
They are consistent with the ones stated by \citet{camilo}.
However, the published uncertainties appeared to be considerably smaller 
than the estimate based on the synthesised beam size and the pulsar S/N.
We thereby conclude that the published pulsar position errors were severely underestimated.

\begin{figure}
 \setlength{\unitlength}{1mm}
  \begin{center}
    \begin{picture}(90,117)(0,0)
      \put (7,60){\includegraphics[scale=0.43, clip]{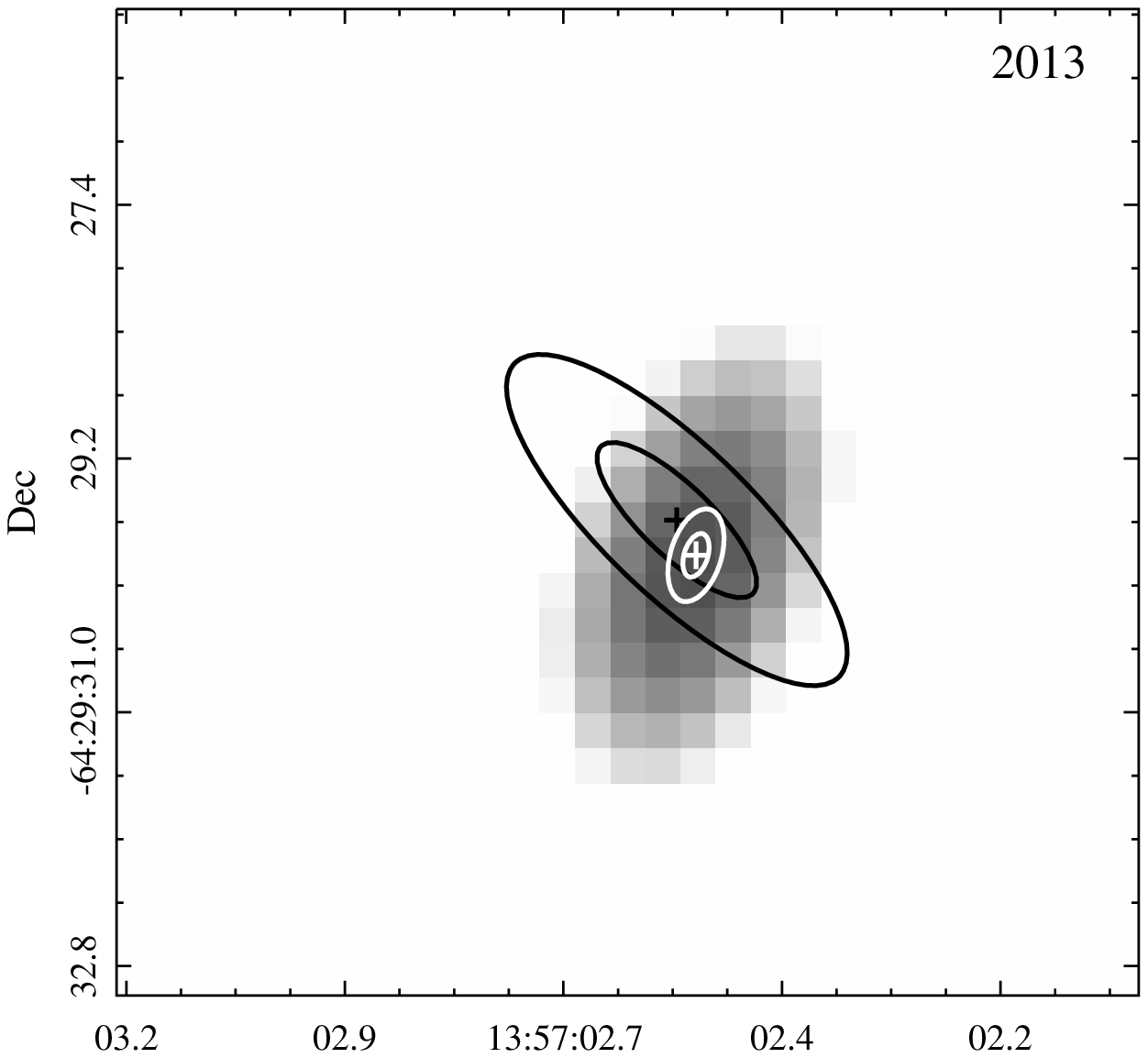}}
      \put  (6,4){\includegraphics[scale=0.45, clip]{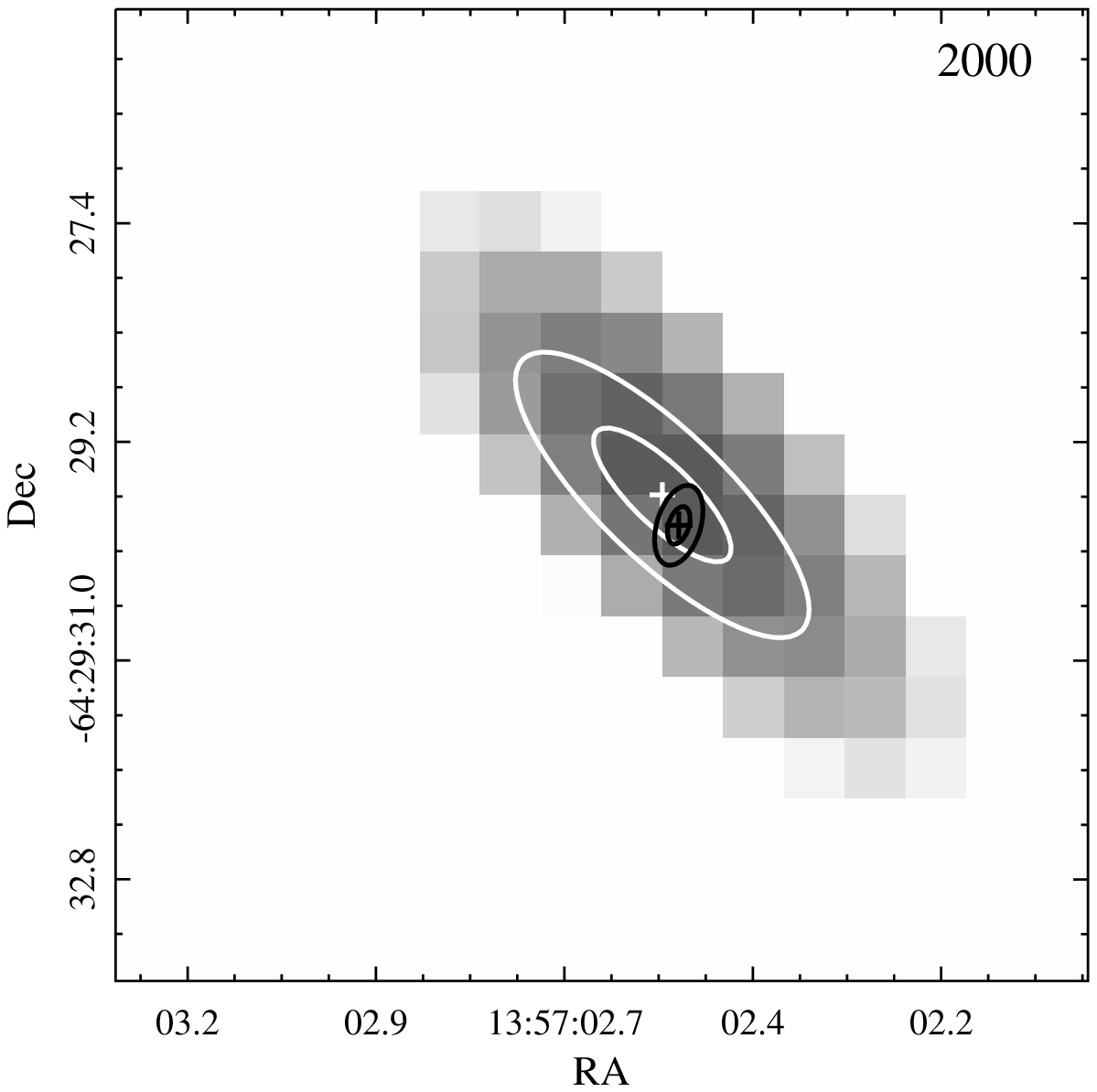}}
 \end{picture}
  \end{center}
  \caption{$8\arcsec\times8\arcsec$ ATCA image fragments
  with  PSR J1357$-$6429 in the centre obtained from the non-self-calibrated data at 2.102 GHz in 2013
  ({\sl top}) and at 1.376 GHz in 2000 ({\sl bottom}) epochs using the same calibrator and  with pixel scales of 0\farcs25 and 0\farcs5,
  respectively.
  For a given epoch the derived pulsar position and the 40 and 90 per cent position uncertainties are shown by  the white cross and ellipses,
  while the black cross and ellipses are related to  the other epoch. The images were used for absolute astrometry. 
  }
\label{fig:abs-pos}
\end{figure}

In Fig.~\ref{fig:abs-pos}, we show the pulsar full-band on-pulse images obtained for two epochs using the same calibrator (1329$-$665).
The pulsar positions for the two observational epochs with the respective error
ellipses at 40 and 90 per cent confidence levels, which account for the systematics mentioned above, are overlaid on the images. 
No significant pulsar shift is visible and
only an upper limit of $\approx1\farcs28$ (90 per cent) can be established.
Accounting for that and a time-base of 12.76~yr between
the two ATCA observations, the 90 per cent upper limit 
on the pulsar proper motion is $\mu \la100$~mas~yr$^{-1}$.

\subsubsection{Relative astrometry}

\begin{figure}
 \setlength{\unitlength}{1mm}
 \begin{center}
  \includegraphics[scale=0.4,clip]{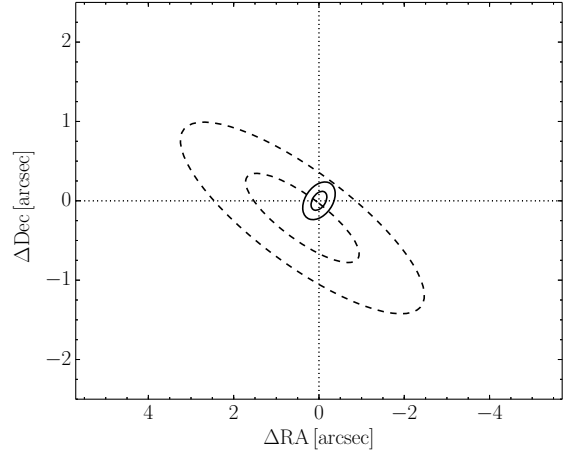}
  \end{center}
  \caption{40 and 90 per cent pulsar position uncertainty ellipses after relative astrometry.
  The 2013 and 2000 positions are shown by solid and dashed ellipses, respectively.
  The positions are shown relative to the mean pulsar position at the 2013 epoch. 
  The offsets differ from those presented in Fig.~\ref{fig:abs-pos} for the absolute astrometry.}
\label{fig:rel-pos}
\end{figure}

\begin{figure}
\begin{center}
 \includegraphics[scale=0.4,clip]{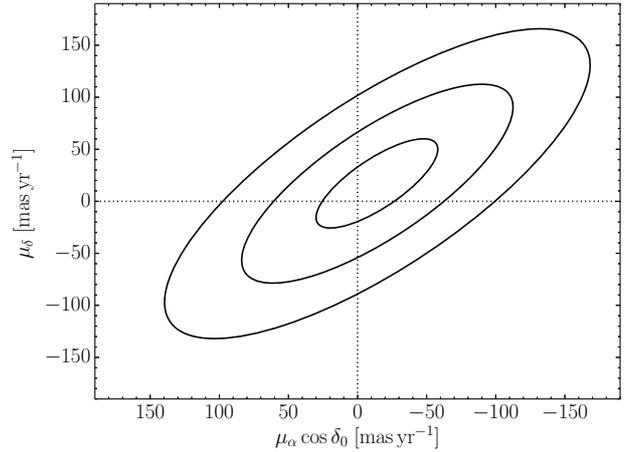}
\end{center}
  \caption{40, 90 and 99.73 per cent confidence regions of the
  pulsar proper motion based on the relative astrometry.
  $\cos{\delta_{0}} \approx 0.43$, where $\delta_{0}$ is the declination of the phase centre for the 2013 epoch data.
  }
  \label{fig:conf-shift}
\end{figure}

Relative astrometry is generally considered as a
more robust tool which 
allows to account for systematic effects which cannot be excluded in advance. 
For instance, comparing 2000 and 2013 positions
of various sources in the field, we found that they exhibit systematic
shifts between the epochs, directed roughly radially from the phase centre and
increasing with the offset from it. 
This stretch could result from bandwidth smearing which was slightly 
different for pre-CABB and CABB data. 
There also can be a small rotation between epochs caused by different ephemeris 
codes used for pre-CABB and CABB correlators.

To increase the image dynamic range and reduce the positional errors, we
used the self-calibrated images for each of the 2013 sub-bands, as well as
for the 2000 epoch. The pulsar S/N was slightly increased as compared to the non-self-calibrated case 
and was $\approx$ 18, 9.9, 11.8 and 8.1 for 
the 1.334, 1.846, 2.358 and 2.780 GHz sub-bands, respectively, while in the 2000 image the S/N remained the same. 
Self-calibration, however, introduced some shift due to an incomplete sky model.
The five images were aligned using a custom-built
routine which accounts for shifts, rotations and stretches. 
For referencing, we used
nine relatively bright point-like sources detected in all images with the signal-to-noise ratio $\ga30$.
They are shown in Fig.~\ref{fig:FOV} and
their positions were determined using {\sl imfit} routine with an accuracy $\la0\farcs07$.
We found the stretch by a factor of 1.0021(3) and a small rotation by 1.2(6) arcmin 
for the 2000 image with respect to the 2013 images.
Neither significant stretch nor rotation between different sub-bands of 2013 observations were found.
After the transformation, positions of the reference sources in
all images became consistent within the uncertainties.
The pulsar proper motion was included in the routine and was
fitted simultaneously with the reference solution.

In Fig.~\ref{fig:rel-pos}, the mean of the pulsar positions in the aligned 2013 sub-bands
is compared with the 2000 position.
We find that the arrangement of the pulsar error ellipses on epochs 2000 and 2013 slightly
differs from that provided by the absolute astrometry (cf., Fig.~\ref{fig:abs-pos}), 
while again no significant 
shift between the epochs is seen. The 40, 90 and 99.73 per cent
confidence regions on the pulsar proper motion $\rm \mu_{\alpha}\cos{\delta_{0}}$ and $\rm \mu_{\delta}$ 
are shown in Fig.~\ref{fig:conf-shift}.
The  90 per cent upper limit on the pulsar proper motion, which is
the radius of the circular region containing 90 per cent probability,
is $\mu < 106$ mas yr$^{-1}$.
This value is compatible, but slightly higher than
the result obtained from the absolute astrometry. 
The reason for this 
is that the positional errors of the 2000 data are in fact underestimated in
the absolute astrometry method since 
the systematic errors are not accessible.
Therefore, we believe that the relative astrometry 
results are more reliable.

\subsection{The pulsar flux densities and spectrum}
\label{sec:flux}
\begin{figure}
  \begin{center}
 \includegraphics[scale=0.4,clip]{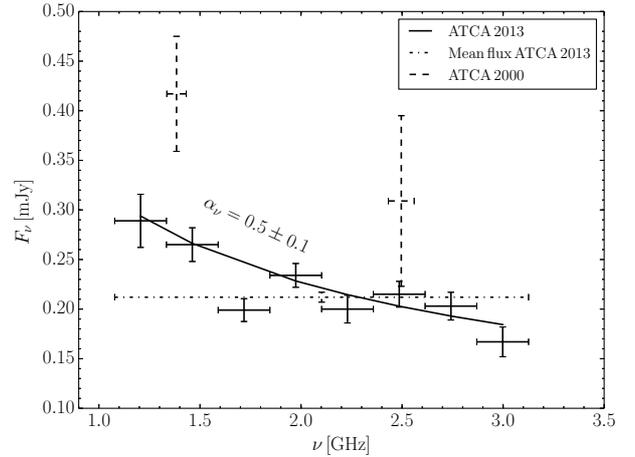}
  \end{center}
  \caption{ATCA spectrum of PSR J1357$-$6429.
  Solid vertical error-bars are the flux densities with 1$\sigma$ 
  uncertainties in sub-bands shown by horizontal solid bars with 256 MHz widths
  regularly spaced within 
  the whole 1.1--3.1 GHz band. The  line represents  the best fit of the measured fluxes by the power
  law $F_{\nu} \propto \nu^{-\alpha_{\nu}}$ with a spectral index $\alpha_{\nu}$ shown in the plot.
  Mean  flux densities   in the whole 2.102 GHz band measured without division 
  by sub-bands and in 1.376 and 2.496 GHz bands of 2000
  are shown by dot-dashed and dashed error-bars  for comparison. The systematic flux calibration 
  uncertainties of $\la$2 per cent are negligible in this scale.
  }
\label{fig:spec}
\end{figure}

To measure the pulsar flux, we used self-calibrated on-pulse visibilities.
To study the spectrum in detail, we additionally split each 512 MHz sub-band
in half. In order to exclude possible errors introduced by the
non-linear cleaning algorithms, we measured the flux directly from the visibility data
using the MIRIAD {\sl uvfit} routine. The pulsar position, however, was fixed at the values
obtained with the {\sl imfit} on the corresponding clean images.
The measured fluxes are shown in Fig.~\ref{fig:spec} by solid error-bar crosses. The mean flux over the full 2013 band is $212\pm5$~$\mu$Jy.
The spectral energy distribution shows a noticeable flux depression at $\approx1.7$ GHz.
The resulting spectrum excluding this point can be fitted
by a single power-law with a spectral index
$\alpha_{\nu}$~= $0.5\pm0.1$ (reduced $\chi^2\approx 0.8$). 
The depression significance in respect of the fit is $\approx3\sigma$. 
The inclusion of this point makes the fit unacceptable (reduced $\chi^2\approx 2.3$).
The depression can hardly be explained by any interstellar absorption at the pulsar line of sight, e.g. owing to 
OH 1.665 and 1.667 GHz lines. There are only 3 pulsars, where weak OH absorption features with FWHM of 1.5$-$3 km s$^{-1}$ are observed 
\citep[e.g.,][and references therein]{minter}. Such features cannot lead to the observed flux depression in our case. 
In addition, we inspected background objects and found a similar 
feature at 1.7 GHz in spectra of the  sources located within $\approx$2\farcm5 of the phase centre, while    
the sources located at larger angular distances and  the calibrators do not show this feature. 
Based on this we suppose that the feature is a systematic   
artifact and  exclude it from the pulsar spectral fit\footnote{ 
According to the operation team, the flux depression can be caused by a correlator issue in the pulsar binning mode. 
}.

The pulsar fluxes on the 2000 data were measured using the full-band self-calibrated
1.376 GHz and 2.496 GHz visibilities. 
The respective fluxes of $417\pm58$~$\mu$Jy and $309\pm86$~$\mu$Jy are shown in Fig.~\ref{fig:spec} by dashed error-bar crosses.
Given the large uncertainties of the 2000 observations, it is difficult to make
any conclusion about the spectral index. However, the 2000 flux values appear to be larger than
those of the 2013 data, at least for the lower frequency. The pulsar flux at 1.376 GHz in the 2000 data differs 
from that in  2013  by about 2.5$\sigma$ including 2 per cent flux calibration 
uncertainties\footnote{ See, e.g., http://www.narrabri.atnf.csiro.au/observing/users guide/html/chunked/ch02s02.html}. The difference  
significance is thus about 99 per cent. 

It is possible that the flux variability can be attributed to the long-term refractive scintillation \citep[e.g.,][]{lorimerkramer}.
Using the galactic electron density model NE2001 of \citet{cordeslazio} we estimated the field coherence scale $s_0\approx 4900$~km,
the Fresnel scale $l_{\rm F}\approx 1.6\times 10^6$~km and the refractive scale 
$l_{\rm R}=l_{\rm F}^2/s_0\approx 5.2\times10^8$~km at 1.376~GHz for the pulsar line of sight. Given that $l_{\rm F} \ll l_{\rm R}$,    
the intensity modulation index due to the refractive scintillation can be estimated as $m_{\rm R}=(s_0/l_{\rm R})^{1/6}\approx 0.15$ \citep{lorimerkramer}. 
This value is consistent with the observed  1.376~GHz flux modulation between the two epochs $(F_{2000}-F_{2013})/(F_{2000}+F_{2013})=0.16\pm0.08$. 
For the Kolmogorov turbulence spectrum, the modulation index scales with the frequency as $m_R\propto \nu^{0.56}$ and at 2.496 GHz is larger by a factor 
of 1.4 than at 1.376~GHz. This is not excluded by the data.
The timescale of the flux variability due to the refractive scintillation $\Delta t_{\rm R} = l_{\rm R}/v_{tr}$ is inversely proportional 
to the pulsar transverse velocity $v_{tr}$. 
Using the derived upper limit on the pulsar proper motion we estimated the lower limit $\Delta t_{\rm R} \gtrsim 6$~days, 
which is significantly larger than the duration of our observations.

\begin{figure*}
 \begin{center}
   \includegraphics[scale=0.5,clip]{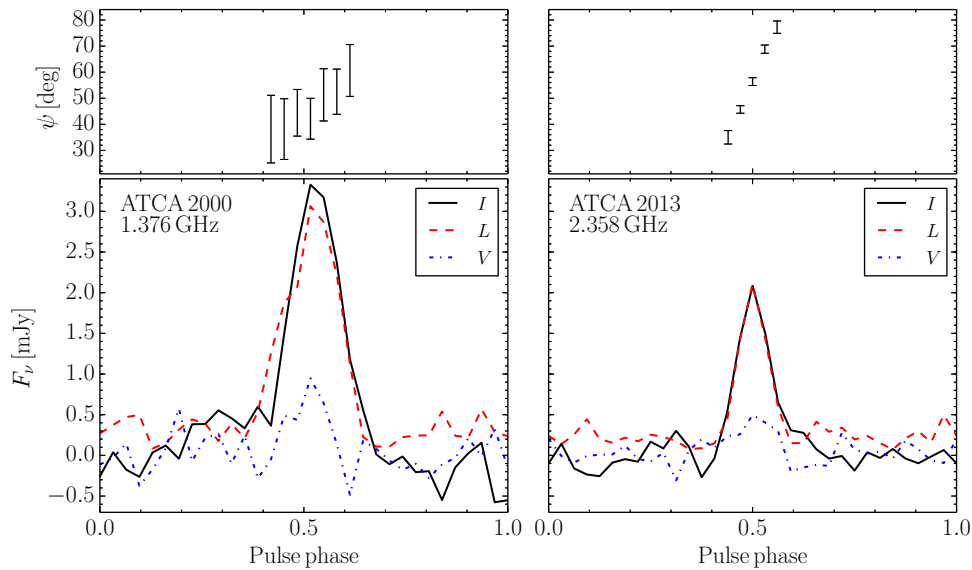}
 \end{center}
 \caption{Pulse and polarisation profiles ({\sl bottom}), and linear polarisation 
 angle $\psi$ ({\sl top}) for PSR J1357$-$6429
 obtained with the ATCA at 1.376 GHz ({\sl left}) and  2.358 GHz ({\sl right}). One full period is shown. 
 Solid, dashed and dot-dashed lines show the total intensity $I$, linearly polarised component $L=(Q^2+U^2)^{1/2}$
 and circular polarisation $V$, respectively. $\psi$ is corrected for RM and shown for on-pulse phases. 
 Phase of 0.5 is placed at the flux density peak. 
 }
 \label{fig:prof}
\end{figure*}

\subsection{The pulse profile and polarisation}

We studied the pulse and polarisation profiles of the 2013 epoch
in the same eight 256~MHz sub-bands. The phase-resolved $I$, $Q$, $U$
and $V$ Stokes parameters
were extracted using the {\tt MIRIAD} {\sl psrplt} routine. The
pulsar emission is observable in five adjacent phase bins.
We found no significant profile width and  polarisation  changes
across the whole 2.102 GHz band. As an example, the resulting
pulse profiles for the Stokes $I$, the circular polarisation $V$ and the
linearly polarised component $L=(Q^2+U^2)^{1/2}$
in the 2.358 GHz sub-band are shown in the lower right panel of Fig.~\ref{fig:prof}. 
The intensity profile shows a relatively wide single
component. The pulse FWHM of $32\degs\pm3\degs$ obtained using a Gaussian profile fit, 
virtually 100 per cent of the linear polarisation and small amount of circular polarisation 
are fully consistent with the higher
time and spectral resolution 1.4 GHz data obtained with the Parkes telescope
\citep{johnston,rookyard2015MNRAS}. 
For completeness, in the lower-left
panel in Fig.~\ref{fig:prof} we show the same profiles for the
1.376~GHz band of the 2000 data set. These data demonstrate
somewhat wider intensity profile with FWHM $\approx 43\degs\pm4\degs$.
The broader pulse profile observed in the 2000 data
can partially result from the DM smearing 
of  $\approx$~9\degs~due to a twice as low spectral 
resolution with the channel width of $\approx$ 9 MHz. 
The smearing  in the 2013 data at the same frequency is less 
significant, $\approx$~4\degs, and it does not affect the  pulse 
profiles at higher frequencies. Accounting for the DM smearing and 
the profile fit uncertainties, 
the 2000 and 2013 intrinsic pulse  widths appear to be consistent.  

\begin{figure}
  \begin{center}
    \includegraphics[scale=0.4, clip]{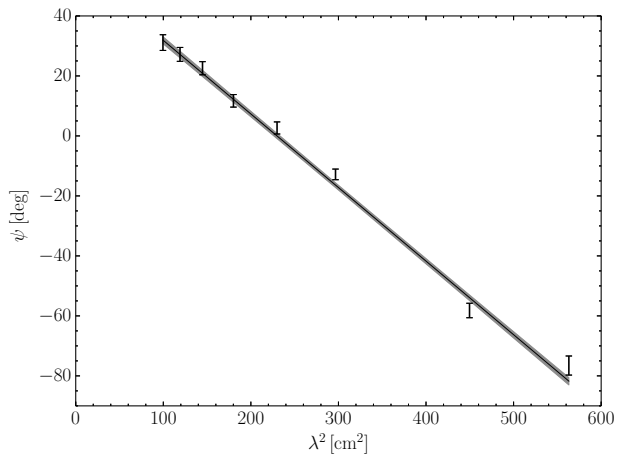}
  \end{center}
  \caption{Polarization angle $\psi$  
  against the square of the observing
  wavelength $\lambda$.
  The grey-hatched region shows 68 per cent linear fit uncertainty. 
  }
\label{fig:RM}
\end{figure}

The large bandwidth of the 2013 data allowed us to infer the
rotation measure (RM) along the pulsar line of sight utilising the
$\lambda^2$ dependence of the linear polarisation position angle
$\psi$. The latter values and their errors were determined from
the Stokes parameters Q and U measured in the five on-pulse phase bins
with the {\it uvfit} routine (as described in Sec.~\ref{sec:flux}) for each of
the eight spectral sub-bands. The linear fit to the observed
$\psi(\lambda^2)$ dependence was performed in each phase bin in
order to check for the possible RM variation with phase \citep{noutsos2009MNRAS}. 
The resulting RM values were found to be consistent
within uncertainties, therefore the global fit for $\psi(\lambda^2)$ dependence was performed 
with RM value tied between all five bins. The
initial regression was not good, with $\chi^2=1.79$ for
34 degrees of freedom, indicating that either position angle
errors were underestimated or there exists a non-trivial spectral
behaviour of the position angle. Indeed, the data for two lowest
frequency bands in our set likely show a shallower swing of $\psi$
than in the other six sub-bands. Since the reason  
for such behaviour is unclear, we  
conservatively increased the errors of the
measured $\psi$ values to make $\chi^2$ of the global
fit equal one. The resulting RM of $-43\pm{1}$
rad m$^{-2}$ is smaller, but consistent with $-47\pm{2}$ rad
m$^{-2}$ estimated from the Parkes Telescope 1.4 GHz data obtained
with a narrow bandwidth \citep{johnston}. We note also, that the
RM fit for the six upper frequencies is successful without any error
renormalisation, however giving smaller RM of $-37\pm 1$ rad
m$^{-2}$. Therefore, we can not exclude the spectral variations of the RM, as observed for some pulsars \citep{dai2015}.
The illustration of the $\psi(\lambda^2)$ dependence for
the central phase bin is shown in Fig.~\ref{fig:RM}. The
solid line with grey-hatched region in Fig.~\ref{fig:RM} show the
best-fit linear regression with 68 per cent uncertainty region.

The phase-dependence of the position angle in the five
on-pulse bins, corrected for the rotation measure, is shown in the
top-right panel in Fig.~\ref{fig:prof}.
This dependence is clearly linear
with the slope  $C = 0.96 \pm 0.03$. 
As expected, the results for 1.376~GHz data of  2000
observations are consistent with the 2013 measurements
(Fig.~\ref{fig:prof}, top-left panel). They are also
consistent with the higher time resolution data 
obtained with the Parkes telescope \citep{rookyard2015MNRAS}. 
The $\psi$-phase dependence demonstrates  almost no curvature.
Therefore, virtually any pulsar emission
geometry can be fitted to the data, according to the classical
rotation vector model \citep[RVM;][]{radhakrishnan}. 
Using a more detailed $\psi$ curve, \citet{rookyard2015MNRAS} obtained loose constraints on
the angle between the magnetic and the rotation axes $\alpha$.  
However, the authors favoured the almost aligned rotator
with $\alpha\approx 7^{\circ}$.

While RVM does not provide robust constrains on the pulsar
emission geometry from our data, it is nevertheless possible to invoke
additional arguments basing on the observed pulse profile width
and the statistical data on other pulsars. \citet{gil}
showed that the half-opening angle of the pulsar beam $\rho_{q}$ (at a
certain relative intensity level $q$) can be related to the pulse
width  $W_{q}$ at the same level via the expression 
\begin{equation}\label{eq:rho}
  \cos\rho_{q}=\cos\alpha\cos(\alpha+\beta)+\sin\alpha\sin(\alpha+\beta)
  \cos(W_{q}/2),
\end{equation}
where $\beta$ is the angle of the closest impact of
the line of sight to the magnetic axis.
In addition, the slope of the positional angle curve at the
inflection point is related to the angles $\alpha$ and
$\beta$ via the relation
\begin{equation}\label{eq:C}
   \sin\alpha= C\sin\beta
\end{equation}
\citep{komes}. Utilising these equations and the value
for the pulse FWHM of $W_{50}=32\degs$, we can find the angles
$\alpha$, $\beta$ and $\rho_{50}$ for any line of sight position
$\zeta=\alpha+\beta$.  The resulting solution varies in the broad
range, however the ratio $|\beta|/\rho_{50}$, i.e. the value of
the impact parameter in units of the emission cone radius, is tied
in the narrow range of $0.93$--$1$ for all $\zeta$. Similar
solutions with $W_{10}=70^\circ$ give
$|\beta|/\rho_{10} \in (0.77,1)$. This suggests that the line of
sight passes at the edge of the beam with the rotation of the
star. One can try to extract the  real value of $\rho$ from the
statistical data on other pulsars. The diagram of the pulse
widths $vs$ period has a clear low boundary which, in case of the
core component, is thought to represent the value of
$\rho$ for orthogonal rotators ($\alpha\approx\pi/2$) with low
impact parameters ($\beta\approx 0$) \citep{rankin90, macies2}. 
In this case, the main source of the difference of the
observed width from the minimal value is described by 
\begin{equation}\label{eq:rankin}
W_{\rm
obs}=\frac{W_{\rm \min}}{\sin\alpha}
\end{equation}
\citep{rankin90, macies2}. 
Using the low boundary from \citet{macies2}, we can
estimate the inclination angle $\alpha$ to be about $10^\circ$ on
the basis of the $W$ measurements alone. Note, that according to
\citet{macies2}, this method can overestimate $\alpha$ in the
close-to-aligned cases (low $\alpha$). However, it is more likely
that the single pulse of PSR J1357$-$6429 is the cut of the cone component at its
side because of the observed high linear and low circular
polarisations \citep{lyneman}. \citet{macies3} argue that the estimate (\ref{eq:rankin})  
is still applicable for the conal profiles, provided impact
angle $\beta$ is small. It is not likely in the case of PSR J1357$-$6429. 
Since the relative contributions of the profile broadening with
$\alpha$ and narrowing with $\beta$ are not known in advance, we
employed the method suggested by \citet{malov_geo}. 
The authors
recommend to use the mean value of the observed pulse widths $W$ in the sample of pulsars with similar periods as
the conservative estimate for twice the value of $\rho$.
The expression (9) from \citet{malov_geo} yields $\rho_{10}\approx 13^\circ$ for
the PSR J1357$-$6429 period. From the Eqs.~(\ref{eq:rho}) and (\ref{eq:C})
we again obtained $\alpha\approx 10^\circ$, in accordance with the
simple estimate from Eq.~(\ref{eq:rankin}) and the 
favoured solution of \citet{rookyard2015MNRAS}.

\section{Summary}
\label{sec4}

The new ATCA observations of the PSR J1357$-$6429 allowed us to detect the pulsar in the 1.1--3.1 GHz band,
to measure its accurate position, the intensity and polarisation pulse profiles, 
to derive the RM in a wide frequency range
and to study the pulsar spectrum.
Comparing our results with the archival ATCA 2000 
data at 1.376 and 2.496 GHz, we examined variation of the pulsar flux, pulse profile
and polarisation, and constrained the pulsar proper motion.

Based on the ATCA 2013 data, the pulsar has a single
pulse component with a constant pulse width of about $32^{\circ}$ over the whole 
1.078--3.126 GHz frequency range. 
The radiation has a high, almost 100 per cent, linear polarisation, which is
typical for young pulsars with $\dot E \ga 10^{34}$~erg~s$^{-1}$ \citep{weltevrede}.
We do not detect spectral variation of the linear polarisation degree.
The pulse profile shape and polarisation properties are in agreement with 
the Parkes telescope 1.4 GHz observations \citep{camilo, johnston, lemo, rookyard2015MNRAS} 
and the ATCA 1.376 GHz observations 
in 2000. An apparent increase of the observed  pulse width in the 2000 data set 
is mainly due to the DM smearing caused by the worse spectral resolution. 
The pulsar spectral energy distribution extracted from the 2013 data demonstrates 
a shallow spectral
index of about 0.5. 
The ATCA 2000 flux values appear to be larger than
those of the 2013 data, at least for the lower frequency. 
This variability can result from the effects of the refractive interstellar scintillation on the 
timescales significantly larger than the duration of our observations. Detailed monitoring campaign is necessary to check this
\citep[e.g.,][]{bhat1999}.

The rotation measure estimate of $-43\pm{1}$ rad m$^{-2}$
is compatible within 2$\sigma$ with the previous narrow band
measurements but appears to be more reliable since it is obtained
using a wider frequency range. Based on the derived RM, we
estimated the mean Galactic magnetic field along  the pulsar line
of sight $\bar{B} = 1.23\times{10^{-6}}$ RM/DM = $-0.41\pm{0.01}$
$\mu$G. This  is close to the values  for most of other pulsars
from the ATNF
catalogue\footnote{http://www.atnf.csiro.au/research/pulsar/psrcat/,
\citet{manchester}} located within a $2\degs$ circle around the
pulsar position.

\begin{figure}
  \begin{center}
    \includegraphics[scale=0.46,clip]{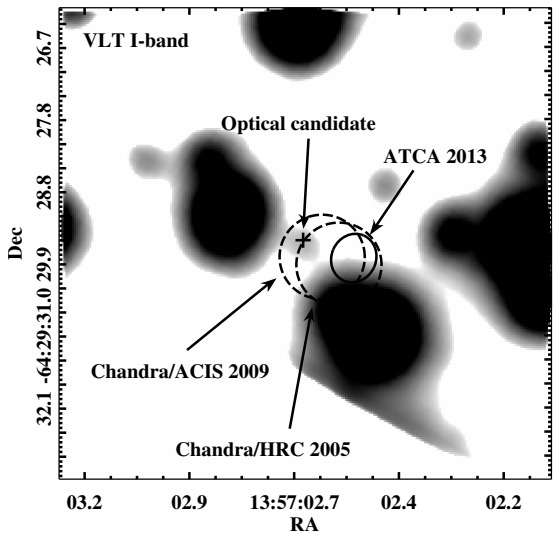}
  \end{center}
  \caption{Fragment of the pulsar field imaged with VLT/FORS2 in $I$-band \citep{danilenko}. 
  The pulsar optical counterpart candidate position is shown by the cross.
  \textit{Chandra} and ATCA pulsar positions are shown by 1$\sigma$ error ellipses.
  These ellipses also account for the optical astrometric referencing uncertainty of 0\farcs2.
  }
\label{fig:VLT}
\end{figure}

On the basis of the absolute and relative astrometry methods,
we did not detect the proper motion of the pulsar.
We estimated a 90 per cent upper limit on it of about 100 mas~yr$^{-1}$ which 
corresponds to the pulsar transverse velocity $v_{tr}\la1200$ km s$^{-1}$ if
the DM distance of 2.5 kpc is adopted.
This does not contradict the value 
of $\sim$~650 km~s$^{-1}$ suggested by \citet{abram} based on 
the offset of the pulsar from the centre of the extended source HESS J1356$-$645, 
possibly associated with the pulsar plerion.
The reason of the discrepancy with the extremely high pulsar transverse velocity suggested by 
\citet{mignani} and \citet{danilenko} is
a strong underestimation of the uncertainties of the pulsar radio coordinates derived from the
ATCA 2000 observations by \citet{camilo}.
Our analysis shows that further radio observations 
with at least the same spatial resolution as provided by the ATCA 2013 data,
will allow one to constrain the pulsar proper
motion shift with accuracy $\la0\farcs15$.

The new ATCA position of the pulsar can be compared with the position of the
suggested candidate to the pulsar optical counterpart \citep{danilenko}.
The ATCA 68 per cent error ellipse is 
overlaid on the VLT $I$-band image adapted from \citet{danilenko} and shown in Fig.~\ref{fig:VLT}.
The pulsar X-ray positions obtained with \textit{Chandra} are also presented.
Given a higher accuracy of the ATCA pulsar position as compared with the X-ray positions,
we conclude that the optical candidate can be discarded as the pulsar counterpart 
at a 99 per cent significance.
A relatively bright star is located in $\sim1\arcsec$ southwards the new radio position of the pulsar,
and further searching for its faint optical counterpart
is only possible with a high spatial resolution imaging provided either 
by the HST or ground-based optical telescopes with adaptive optics systems.

\section*{Acknowledgements}
We are grateful to the referee Adam Deller for his comments which prompted us to reconsider certain parts of the manuscript.
Authors thank E. B. Nikitina, I. F. Malov, F. Camilo, Serge Balashev and D. E. Alvarez-Castillo for useful discussions. 
The work was supported by the Russian Science Foundation, grant 14-12-00316.
The Australia Telescope Compact Array
is part of the Australia Telescope National Facility which is funded
by the Commonwealth of Australia for operation as a National Facility managed by CSIRO.
%%%%%%%%%%%%%%%%%%%%%%%%%%%%%%%%%%%%%%%%%%%%%%%%%%



\bibliographystyle{mnras}
\bibliography{ref0357} % if your bibtex file is called example.bib


%%%%%%%%%%%%%%%%%%%%%%%%%%%%%%%%%%%%%%%%%%%%%%%%%%

%%%%%%%%%%%%%%%%% APPENDICES %%%%%%%%%%%%%%%%%%%%%

%%%%%%%%%%%%%%%%%%%%%%%%%%%%%%%%%%%%%%%%%%%%%%%%%%

% Don't change these lines
\bsp	% typesetting comment
\label{lastpage}
\end{document}